\documentclass[aps,prb,twocolumn,superscriptaddress, noeprint]{revtex4-2}
\usepackage{bbold}
\usepackage{comment}
\usepackage[colorlinks, citecolor=red]{hyperref}
\usepackage{hyperref}
\usepackage[utf8]{inputenc}
\usepackage[T1]{fontenc}    %
\usepackage[english]{babel}
\usepackage[pdftex]{graphicx}
\usepackage{amsmath,amssymb,amsthm,bbm, mathtools} %
\usepackage{mathrsfs}
\usepackage{tikz}   %
\usepackage[normalem]{ulem}

\definecolor{green(html/cssgreen)}{rgb}{0.0, 0.5, 0.0}

 \date{\today}
 
\DeclareMathSymbol{\shortminus}{\mathbin}{AMSa}{"39}
\renewcommand{\vec}{\boldsymbol}

\begin{document}

\title{Coupled superconducting spin qubits with spin-orbit interaction}

\author{Maria Spethmann}
\thanks{These two authors contributed equally to this work} 
\affiliation
{Department of Physics, University of Basel, 4056 Basel, Switzerland}
\author{Xian-Peng Zhang}
\thanks{These two authors contributed equally to this work}
\affiliation
{Department of Physics, University of Basel, 4056 Basel, Switzerland}
\author{Jelena Klinovaja}
\affiliation
{Department of Physics, University of Basel, 4056 Basel, Switzerland}
\author{Daniel Loss}
\affiliation
{Department of Physics, University of Basel, 4056 Basel, Switzerland}

\begin{abstract}
Superconducting spin qubits, also known as Andreev spin qubits, promise to combine the benefits of superconducting qubits and spin qubits defined in quantum dots. While most approaches to control these qubits rely on controlling the spin degree of freedom via the supercurrent, superconducting spin qubits can also be coupled to each other via the superconductor to implement two-qubit quantum gates. 
We theoretically investigate the interaction between superconducting spin qubits in the weak tunneling regime and concentrate on the effect of spin-orbit interaction (SOI), which can be large in semiconductor-based quantum dots and thereby offers an additional tuning parameter for quantum gates.
We find analytically that the effective interaction between two superconducting spin qubits consists of Ising, Heisenberg, and Dzyaloshinskii-Moriya interactions and can be tuned by 
the superconducting phase difference, the tunnel barrier strength, or the SOI parameters. The Josephson current becomes dependent on SOI and spin orientations.
We demonstrate that this interaction can be used for fast controlled phase-flip gates with a fidelity >99.99\%. 
We propose a scalable network of superconducting spin qubits which is suitable for implementing the surface code.
\end{abstract}

\maketitle

\section{Introduction}
Superconducting qubits \cite{nakamura1999coherent,chiorescu2003coherent,devoret2013superconducting} and spin qubits \cite{Loss1997,hanson2007spins, Kloeffel2013, awschalom2013quantum,burkard2021semiconductor,xue2022quantum} are promising candidates to build a quantum computer, as they can be scaled up using the fabrication techniques of the semiconductor industry. While superconducting qubits are experimentally more advanced, they typically have a macroscopic size ($\sim 0.1$ mm \cite{krinner2019engineering}). %
Spin qubits, on the other hand, show long coherence times \cite{veldhorst2014} and are very compact ($\sim 50$ nm \cite{xue2022quantum}), but the small size poses a challenge to incorporate all the gates needed to control the spins individually.

Recently,  superconducting spin qubits \cite{chtchelkatchev2003andreev, padurariu2010theoretical,Park2017,hays2021coherent}, also known as Andreev spin qubits, have become a growing research field which combines the concepts of superconducting qubits and spin qubits. For this, a quantum dot is placed in a Josephson junction where the normal part hosts the subgap bound states, the Andreev bound states \cite{meng2009self,Prada2020}. The odd-occupancy Andreev bound states define a spin qubit that couples to the  supercurrent \cite{chtchelkatchev2003andreev}.

While superconducting spin qubits can be coupled to each other through the superconducting current
(``inductive coupling'') \cite{chtchelkatchev2003andreev}, another way of coupling is by effective wave-function overlap. When two quantum dots couple to the same superconductor at a distance smaller than the superconducting coherence length ($\sim 0.7$~$\mu$m \cite{mayer2019superconducting}), spins located on these dots interact isotropically via Cooper pairs that split 
~\cite{Choi2000,Choi2001,LeHur2006,Hassler2015,GonzalezRosado2021}.
This length scale is significantly longer than the direct exchange of regular spin qubits ($\sim 50$ nm \cite{xue2022quantum}), which makes it easier to add control and readout components.
Furthermore, coupling through overlap \cite{Kornich2019} is tunable by the superconducting phase difference \cite{Choi2000} and may allow for a more complex qubit connectivity, as discussed below.
Importantly, in contrast to inductive coupling, which is quadratic in spin-orbit interaction (SOI) strength \cite{padurariu2010theoretical}, our spin coupling does not vanish in the absence of the SOI. 
Experimentally, directly overlapping Andreev bound states have been demonstrated in experiments on so-called Andreev molecules \cite{higginbotham2015parity,junger2021intermediate,kurtossy2021andreev,matsuo2021observation}, where the coupling is so strong that two Andreev bound states hybridize \cite{Scherubl2019, Pillet2019}. The coupling mechanism is based on crossed Andreev processes, very similar to Cooper pair splitters \cite{recher2001andreev,Hofstetter2011, Baba2018, Ueda2019, Ranni2021, Kanne2021, kurtossy2022}.
These recent experiments provide strong motivation for the setup proposed here.

The main aspect we wish to analyze here is how SOI \cite{dell2007josephson} affects the exchange interaction between superconducting spin qubits. SOI is a desired property of the quantum dot: superconducting spin qubits in their initial proposal require SOI to control the spin degree of freedom via the supercurrent \cite{chtchelkatchev2003andreev,padurariu2010theoretical, hays2021coherent,tosi2019spin}. But also regular spin qubits need SOI to perform single-qubit rotations via electric dipole spin resonance (EDSR)
\cite{golovach2006,Kloeffel2013}, which allows for all-electrical control. In any case, SOI will likely be present in semiconductor devices and can even be tuned with gates \cite{nitta1997gate,  roulleau2010, kloeffel2011strong, vanweperen2015, Manchon2015, scherubl2016, brauns2016,takase2017highly,Froning2021, dorsch2021, froning2021ultrafast, adelsberger2022}, adding a further control parameter. 
We here want to answer the following questions: How does the SOI affect the effective interaction between superconducting spin qubits? Can this effect be used to implement efficient quantum gates?

We derive a low-energy spin Hamiltonian and find that the SOI induces some ``twists'' in the otherwise isotropic exchange interaction, which results in not only Heisenberg, Ising, and Dzyaloshinskii-Moriya (DM) interactions \cite{Dzyaloshinskii1958,moriya1960anisotropic} (similar to Refs.~\cite{Imamura2004,Klinovaja2013}), but also effective staggered magnetic fields. By tuning the SOI strength and the superconducting phase difference, we can either undo the twist or tune the exchange interaction to be of pure Ising type, which allows one to realize a fast controlled phase-flip gate for two qubits. We analyze the second possibility in more detail and find that we only need to fine-tune one of the parameters (either the superconducting phase or the SOI strength) to reach a high gate fidelity. Finally, we sketch how to arrange the superconducting spin qubits in a two-dimensional scalable array, suitable for the surface code~\cite{bravyi1998quantum}.

We organize the remainder of this paper as follows. In Sec.~\ref{modelandtheory} we introduce our theoretical model of the quantum dots and the superconductors and derive the spin-spin interaction between the spin qubits. In Sec.~\ref{controlledphase} we show how this interaction can be used to realize a fast controlled phase-flip gate and we calculate the gate fidelity. In Sec.~\ref{scalablenetwork} we outline a scalable design of a qubit network. Section \ref{conclusion} summarizes
our findings. Finally, Appendix \ref{fourththeory} gives the details of the perturbation theory, Appendix \ref{ABflux} shows the results accounting for an Aharonov-Bohm flux, and  Appendix \ref{Josephonsupercurrent} presents the exact expressions of the Josephson supercurrent.

\begin{figure*}[!t]
	\centering
	\includegraphics[width=0.75\linewidth]{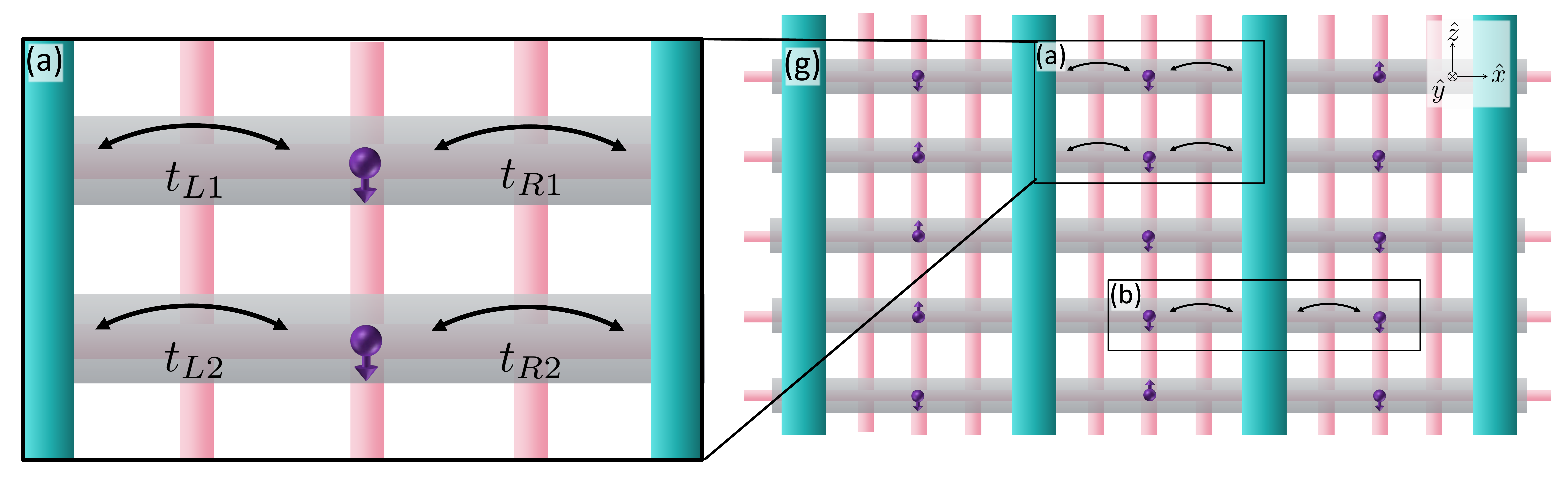}
	\includegraphics[width=0.2\linewidth]{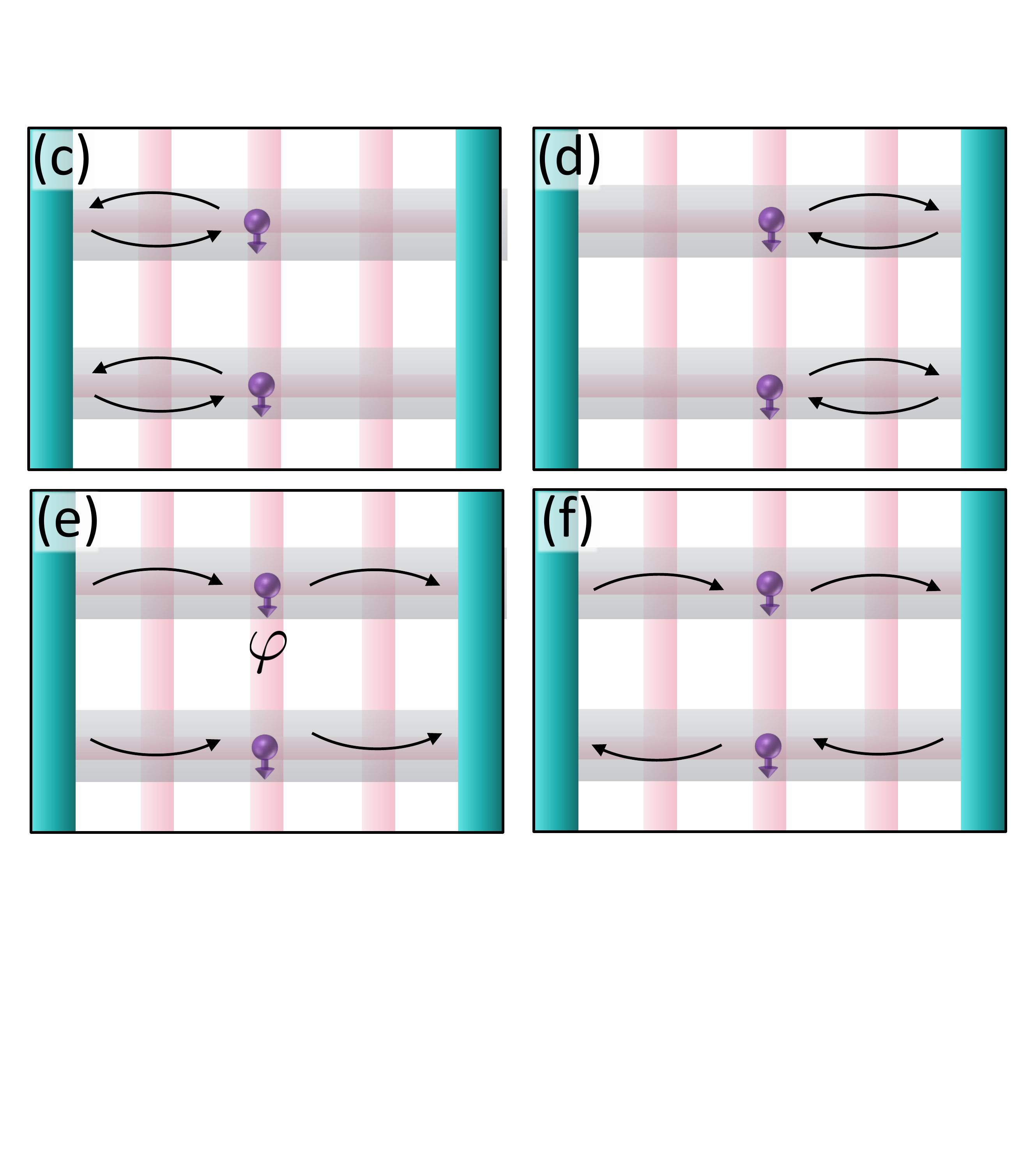}
	\caption{
	(a) Our model describes two spin-1/2 quantum-dot states (violet) that are weakly tunnel coupled ($t_{jn}$) to two superconducting leads (green) such that they form superconducting spin qubits. Due to the presence of SOI, the tunneling is not spin conserving. (b) The superconducting spin qubits can also interact with each other via one superconductor only. (c-f) Virtual tunneling paths that contribute to the spin interaction given in Eq.~\eqref{vfdnakfvm01}. Black arrows show the virtual tunneling paths of the electrons. The contributions from panel (e) depend on the superconducting phase difference $\varphi$. (g) We propose a scalable architecture of superconducting spin qubits. The setup consists of parallel nanowires (gray), orthogonal superconducting stripes (green), and a crossbar design of gates (pink) that form quantum dots in the nanowires and control the strength of tunnel barriers separating them from superconductors. When corresponding tunnel barriers are lowered, the superconducting spin qubits interact pairwise with each other, see panel (a).
	}
	\label{pic01}
\end{figure*}

\section{Model and theory} \label{modelandtheory}
We describe a model of two quantum dots in double nanowires \cite{Kanne2021, kurtossy2021andreev, Vekris2021, Vekris2021Asymmetric, kurtossy2022} that are tunnel coupled to two superconducting leads, as shown in Fig.~\ref{pic01}(a). We consider the limit of weak tunnel coupling, large on-site Coulomb repulsion, and negative dot detuning from the Fermi energy, such that in the ground state the dot is occupied by one electron. This parameter choice describes a superconducting spin qubit \cite{meng2009self}\footnote{The fixed dot filling also suppresses quasiparticle poisoning, thus stabilizing the superconducting spin qubit \cite{karzig2017}}. (Superconducting spin qubits also exist at different parameter regimes and in atomic break junctions \cite{zgirski2011evidence,bretheau2013supercurrent,janvier2015coherent}.)

The total Hamiltonian consists of three parts, $H_{\text{tot}}=H_D+H_L+H_T$. The dot Hamiltonian is 
\begin{equation}\label{eq:Hdot}
H_D= \sum_{ s n}\epsilon d^{\dagger}_{ns}d^{}_{ns} + C_d d^{\dagger}_{n\uparrow}d_{n\uparrow}d^{\dagger}_{n\downarrow}d_{n\downarrow},
\end{equation}
where $d^{\dagger}_{ns}$ creates an electron on the dot $n\in\{1,2\}$  with spin $s\in \{\uparrow,\downarrow\}$ and energy $\epsilon<0$, which is equal on both dots and tuned by electrostatic gates. Further, we have an on-site Coulomb repulsion $C_d$. We assume a small dot size such that we only take into account one dot energy level.

The leads are conventional singlet superconductors and are described by the $s$-wave BCS Hamiltonian:
\begin{equation}\label{eq:Hsc}
H_{L}=\sum_{j\vec{k}s}\xi_k c_{j\vec{k}s}^{\dagger}c^{}_{j\vec{k}s}-\sum_{j\vec{k}}(\Delta_{j} c_{j\vec{k}\uparrow}^{\dagger}c^{\dagger}_{j\,\shortminus\vec{k} \downarrow} + \text{H.c.}).
\end{equation}
Here, $c_{j\vec{k}s}^{\dagger}$ is the creation operator of lead $j\in\{L,R\}$ with spin $s$, wave vector $\vec{k}$, and energy $\xi_k$.
The pairing potential $\Delta_j=\Delta e^{-i\varphi_j}$ is characterized by the superconducting gap $\Delta$ and phase $\varphi_j$. The phase difference between the superconductors is $\varphi=\varphi_L-\varphi_R$.

The leads and the dots are tunnel coupled:
\begin{align} \label{eq:H_T}
H_T&%
=\sum_{jn\vec{k}}t_{jn} c^{\dagger}_{j\vec{k}}U_{jn}d^{}_{n}+\text{H.c.},
\end{align} 
with $d_{n}=\begin{psmallmatrix}d_{n\uparrow}\\d_{n\downarrow}\end{psmallmatrix}$,  $c_{j\vec{k}}=\begin{psmallmatrix}c^{}_{j\vec{k}\uparrow} \\c^{}_{j\vec{k}\downarrow}\end{psmallmatrix}$, positive tunnel amplitude $t_{jn}$, and unitary matrix $U_{jn}\in \text{SU}(2)$. The SOI in semiconducting nanowires 
can be strong \cite{Fasth2007,kloeffel2011strong,Froning2021} and  causes spin rotation \cite{datta1990electronic,Hoffman2017}, which is described by the rotation matrix in spin space $U_{jn}$ \cite{flindt2006spin,dell2007josephson,Li2014}. %
We expect this spin rotation matrix to also describe the spin-orbit mixing in multiorbital dots. Our spin then is the pseudospin of the low-energy states, and on-site spin-orbit mixing just renormalizes our phenomenological rotation matrix \cite{danon2009pauli}.

Next we want to find the effective spin coupling between the dots. Since at least four tunneling processes are needed for spin interaction, we perform a standard fourth-order perturbation expansion in $H_T$ \cite{Choi2000,Schrade2017}. It is valid in the weak-coupling regime $t_{jn},\Gamma_{j}\ll \Delta,|\epsilon|$ \cite{Choi2000,probst2016signatures}, with $\Gamma_{j}=\pi\rho_Ft_{j1}t_{j2}$ where $\rho_F$ is the normal density of states per spin of the leads at the Fermi energy. We also work in the Coulomb blockade regime by assuming that the Coulomb repulsion $C_d$ on the dot is the largest energy scale of the system and forbids double occupation. The influence of finite Coulomb interaction will be discussed elsewhere \cite{zhang2022exchange}. 
We start with a basis transformation of operators defined for the superconducting leads and the second dot:
\begin{align} \label{rotationbasis01}
\begin{psmallmatrix} \Tilde{c}_{R\vec{k}\uparrow}^{} \\ \Tilde{c}_{R\vec{k}\downarrow}^{}
\end{psmallmatrix}&= U^{\dagger}_{R1}\begin{psmallmatrix} c_{R\vec{k}\uparrow}^{} \\ c_{R\vec{k}\downarrow}^{}
\end{psmallmatrix}, \nonumber\\
\begin{psmallmatrix}\Tilde{d}_{2\uparrow}\\ \Tilde{d}_{2\downarrow}\end{psmallmatrix} &= U_{R1}^{\dagger}U^{}_{R2} \begin{psmallmatrix}d_{2\uparrow}\\ d_{2\downarrow}\end{psmallmatrix},\nonumber\\ 
\begin{psmallmatrix} \Tilde{c}_{L\vec{k}\uparrow}^{} \\ \Tilde{c}_{L\vec{k}\downarrow}^{}
\end{psmallmatrix}
&=U^{\dagger}_{R1} U_{R2}^{}  U^{\dagger}_{L2} \begin{psmallmatrix} c_{L\vec{k}\uparrow}^{} \\ c_{L\vec{k}\downarrow}^{}
\end{psmallmatrix}. 
\end{align}
Consequently, in the rotated basis, the electron spin could flip only during the tunneling between dot $1$ and lead $L$, captured by $U=U^{\dagger}_{R1} U_{R2}^{}  U^{\dagger}_{L2}U^{}_{L1}$. Hereafter, we parametrize the total SOI effects by the angle $\alpha$ and by the unitary vector $\mathbf{u}$ defined via $U\equiv e^{i\alpha\mathbf{u}\cdot\mathbf{S}_1}$, which eventually acts on the spin $\mathbf{S}_1$ on dot 1 (see details in Appendix \ref{fourththeory}).

\begin{figure}[t]
\includegraphics[width=0.8\linewidth]{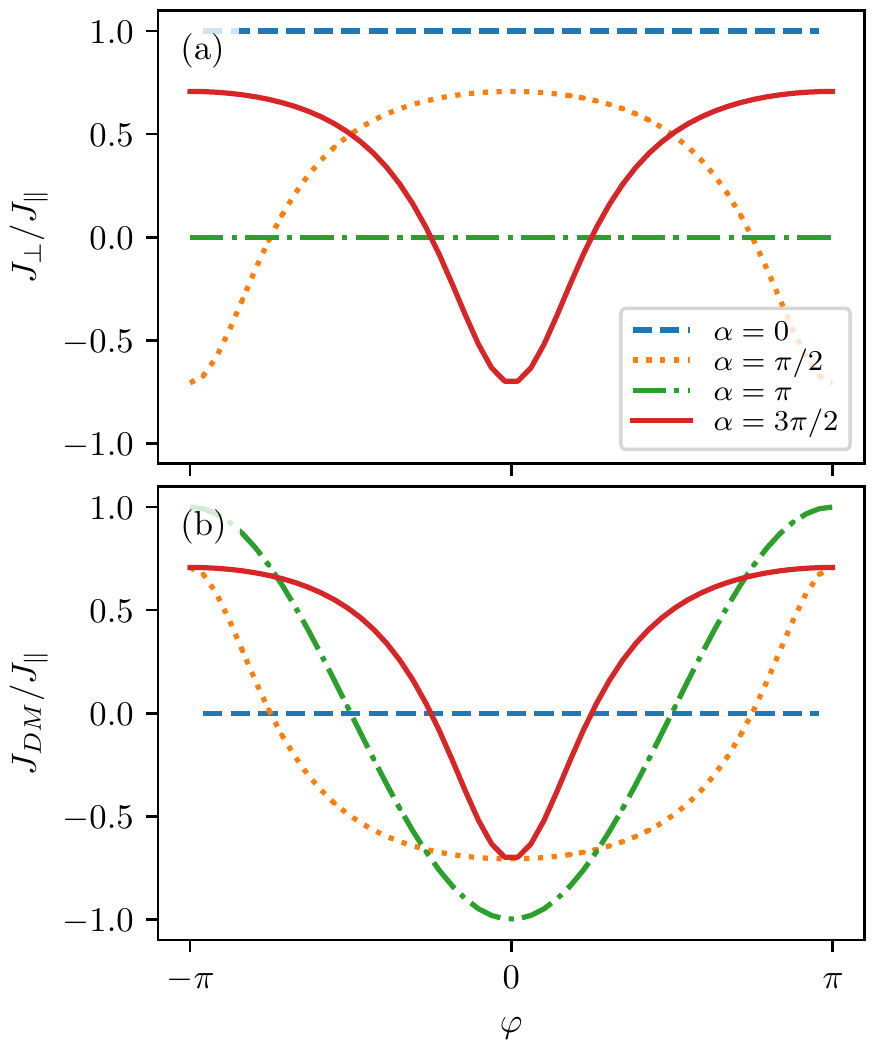}
\caption{The anisotropy of the spin exchange [Eq.~\eqref{vfdnakfvm01}], $J_{\perp}/J_{\Vert}$, and the DM interaction, $J_{DM}/J_{\Vert}$, can be tuned by varying the superconducting phase difference $\varphi$ and the SOI strength $\alpha$. %
If $ \varphi = \pi\pm \alpha/2 $ ($\alpha\neq0$), the interaction is of pure Ising type.
We note that $\alpha$ can be varied while keeping the rotated spin basis on the second dot $\Tilde{S}_2$ unchanged, e.g. by modulating $U_{Ln}$. }
\label{figure2_a}
\end{figure}

We now sum up all virtual paths with four tunneling processes that contribute to a spin interaction between the dots [see Figs.~\ref{pic01}(c)-\ref{pic01}(f) and Appendix \ref{fourththeory}]. Without SOI the spin interaction is isotropic $\propto \mathbf{S}_1\cdot\mathbf{S}_2$ ~\cite{Choi2000}.  Thus the virtual paths in which electrons tunnel only to the  right [see Fig.~\ref{pic01}(d)] or left [see Fig.~\ref{pic01}(c)] lead cause twisted spin exchange \cite{kaplan1983single,PhysRevLett.69.836,PhysRevB.47.174,stepanenko2003spin,Imamura2004,gangadharaiah2008spin,PhysRevB.98.241303}, $\mathbf{S}_1\cdot\tilde{\mathbf{S}}_2$ and $(U^{\dagger}\mathbf{S}_1U^{})\cdot \tilde{\mathbf{S}}_2$, respectively. Here $\tilde{\mathbf{S}}_2=U^{\dagger}_{R}\mathbf{S}^{}_2U^{}_{R}$ is the spin operator in the rotated basis given by Eq.~\eqref{rotationbasis01}, where $U^{}_R=U_{R1}^{\dagger}U^{}_{R2}$. 
We note that one can always eliminate these twisting effects with single-qubit rotations. The ``two-lead paths''  generate an anomalous twisted spin exchange $e^{i\phi}(U^{\dagger}S^{\nu}_1\mathbb{1})\tilde{S}^{\nu}_2+\text{H.c.}$ with $S_n^{\nu}\in\{S_n^o,S_n^x, S_n^y, S_n^z\}$, where $S_n^o=\frac{1}{2}$ and $\phi=\varphi$ [$\phi=0$] correspond to processes shown in Fig.~\ref{pic01}(e) [Fig.~\ref{pic01}(f)]. By ``anomalous'' we mean the left ($U^{\dagger}$) and right ($\mathbb{1}$) twisting matrix of $S^{\nu}_1$ are different, and one cannot eliminate the twisting effect by rotating the spin basis. 
We sum up the ``left-lead,'' ``right-lead,'' and ``two-lead'' paths  and get for $\Gamma_L=\Gamma_R=\Gamma$ and
 $\Delta\gg\vert\epsilon\vert$ \footnote{In this limit, the ``two-lead'' path shown in Fig.~\ref{pic01}(f) can be neglected, see Appendix \ref{fourththeory}.}  the low-energy Hamiltonian 
\begin{align} \label{spinexchangecouplingsmv1}
H&=J\mathbf{S}_1\!\cdot \tilde{\mathbf{S}}_2+J\left(e^{-i\alpha\mathbf{u}\cdot\mathbf{S}_1}\mathbf{S}_1e^{i\alpha\mathbf{u}\cdot\mathbf{S}_1}\right)\cdot \tilde{\mathbf{S}}_2\\
&+ 2J\cos\varphi
\left[\cos\!\left(\textstyle\frac{\alpha}{2}\right)\,\mathbf{S}_1\!\cdot\tilde{\mathbf{S}}_2-\sin\!\left(\textstyle\frac{\alpha}{2}\right)\mathbf{u}\cdot(\mathbf{S}_1\!\times\tilde{\mathbf{S}}_2)\right]\notag \\
&+ J\sin\varphi\sin\left(\textstyle\frac{\alpha}{2}\right)\,\mathbf{u}\cdot\left(\mathbf{S}_1-\tilde{\mathbf{S}}_2\right),\notag
\end{align}
with $J\approx \Gamma^2/\vert \epsilon\vert$.  (In Appendix \ref{ABflux} we present the result without the restrictions $\Gamma_L=\Gamma_R$ and $\Delta \gg\epsilon$.) For $\varphi\neq 0$ or $\pi$, the time-reversal symmetry %
is broken, and anomalous twisted spin-exchange results in  effective staggered magnetic fields [see the third line of Eq.~\eqref{spinexchangecouplingsmv1}]. Here, we dropped a spin-irrelevant function of $\varphi$. From Eq.~\eqref{spinexchangecouplingsmv1} we already see a few important features. For the special case $U\to \mathbb{1}$ and $U_R\to \mathbb{1}$, $H$ reduces into an isotropic Heisenberg-like interaction~\cite{Choi2000}. For $U\to \mathbb{1}$ and $U_R\neq \mathbb{1}$, $H$ reduces to an ordinary twisted spin exchange  \cite{kavokin2004symmetry,Imamura2004,zhu2010electrically} consistent with Ref.~\cite{GonzalezRosado2021}. For $U\neq \mathbb{1}$ the interaction is nontrivial. Defining parallel and orthogonal components of a vector as $x^{\Vert}\!=\!\vec{u}\!\cdot\!\vec{x}$ and $\vec{x}^{\perp}=\vec{x}-x^{\Vert}\vec{u}$, $H$ can be rewritten as
\begin{equation} \label{vfdnakfvm01}
H=h(S^{\Vert}_{1}-\tilde{S}^{\Vert}_{2})+J_{\Vert} S^{\Vert}_{1}\tilde{S}^{\Vert}_{2}+ J_{\perp} \mathbf{S}^{\perp}_{1}\cdot \tilde{\mathbf{S}}^{\perp}_{2}+J_{\text{DM}}(\mathbf{S}_1\times\tilde{\mathbf{S}}_2)^{\Vert},
\end{equation}
where staggered magnetic field and spin-exchange constants, respectively, are given by
\begin{subequations}
\begin{align}
\label{exchangecoupling}
&h=J\sin\varphi\sin\textstyle\frac{\alpha}{2}, \\
&J_{\Vert}= 2J(1 +\cos\varphi\cos\textstyle\frac{\alpha}{2}),
\\ &J_{\perp}=2J\cos\textstyle\frac{\alpha}{2}(\cos\varphi+\cos\frac{\alpha}{2}),\\\ 
&J_{\text{DM}}=-2J\sin\textstyle\frac{\alpha}{2}(\cos\varphi+\cos\frac{\alpha}{2}).
\end{align}
\end{subequations}
 The anisotropy $J_{\Vert}-J_{\perp}$ and the DM interaction $J_{\text{DM}}$ are tuned by the SOI parameter $\alpha$, the lead-dot coupling $\Gamma_j$, and the superconducting phase difference $\varphi$. 
Both $\alpha$ \cite{kloeffel2011strong} and  $\Gamma_j$ \cite{deacon2010tunneling} can be tuned by the electrostatic gates, while a magnetic flux can tune $\varphi$ when the superconductors form a loop \cite{ren2019topological,fornieri2019evidence}. 
To get a better understanding of the spin interaction we plot $J_{\perp}/J_{\Vert}$ and $J_{\text{DM}}/J_{\Vert}$ as a function of $\varphi$ for different $\alpha$ in Figs.~\ref{figure2_a}(a) and \ref{figure2_a}(b). For $\alpha=0$, we reproduce the isotropic interaction limit. The perpendicular component $J_{\perp}/J_{\Vert}$ disappears at $\alpha=\pi$, and one obtains maximum modulation of the DM interaction. Importantly, the interaction takes on pure Ising form for $ \varphi = \pi\pm \alpha/2$ (for $\varphi \neq 0$ or $\pi$). 

Our Hamiltonian $H$ is invariant under the spin rotation around the axis $\mathbf u$, and hence $C=S^{\Vert}_{1}+\tilde{S}^{\Vert}_{2}$ is a conserved quantum number 
such that the setup is also suitable for singlet-triplet qubits \cite{levy2002,Klinovaja2012}.
Furthermore, our effective Hamiltonian is valid for arbitrary strengths of the SOI $\alpha$, in contrast to the inductive coupling in Ref.~\cite{padurariu2010theoretical}, which only considers weak SOI and vanishes 
as $\alpha^2$ $\to 0$.

We note that the SOI induces a spin-dependent phase shift  in the Josephson current, %
$\hat I=\partial_{\varphi}H$, 
given by
\begin{align} \label{1230987123987}
&\hat I=\hat{I}_s\sin\varphi+J\sin\textstyle(\frac{\alpha}{2})(S^{\Vert}_{1}-\tilde{S}^{\Vert}_{2})\cos\varphi,  \\
&\frac{\hat {I}_s}{2J}=\cos\textstyle(\frac{\alpha}{2})\label{454343450908}
\left(\frac{1}{4}-\mathbf{S}_{1}\cdot \tilde{\mathbf{S}}_{2}\right)
+\sin(\frac{\alpha}{2})(\mathbf{S}_1\times\tilde{\mathbf{S}}_2)^{\Vert}.
\end{align}
The $\cos\varphi$ term  enables detection of SOI and the relative spin orientations of the dots, see Appendix \ref{Josephonsupercurrent}.

\section{Controlled phase flip gate} \label{controlledphase}
\begin{figure}
	\centering
	\includegraphics[width=\linewidth]{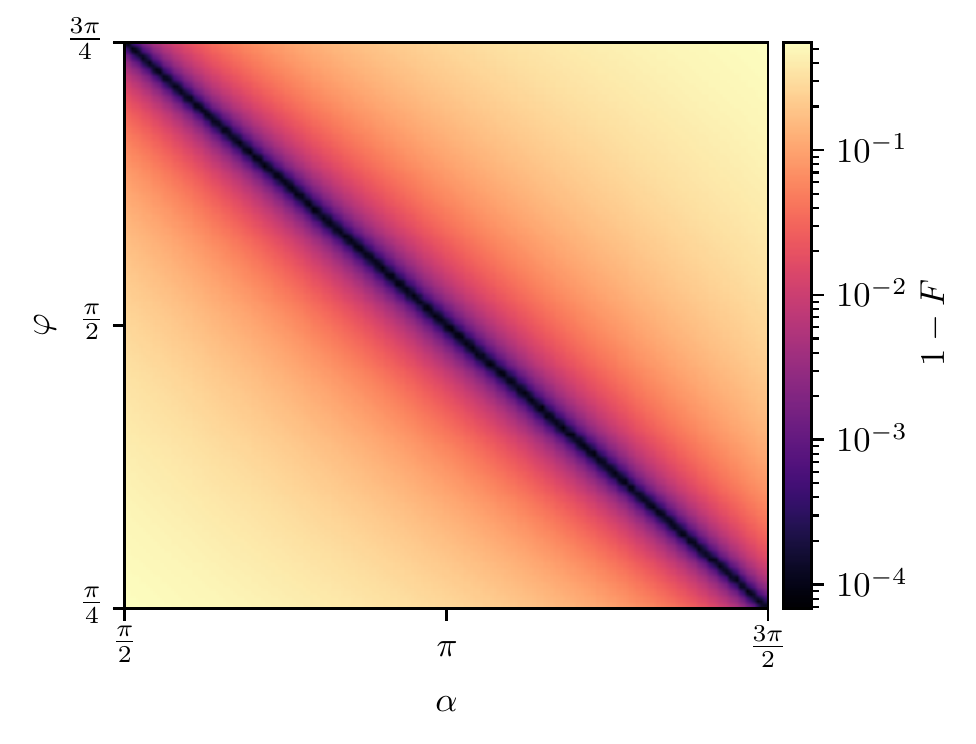}
	\caption{Infidelity $1-F$ of the controlled phase-flip gate $U(\alpha,\varphi)$ for varying  SOI strength  $\alpha$ and superconducting phase difference $\varphi$ [see Eqs.~\eqref{eq:evolution} and \eqref{eq:fidelity}]. Only one parameter needs to be fine-tuned to reach a fidelity $>99.99\%$. We average over statistical Gaussian noise of switching time $t=\frac{\hbar \pi}{J_{\Vert}(\alpha,\varphi)}(1\pm0.003)$, tunnel ratio $\frac{\Gamma_L}{\Gamma_R}=1\pm0.005$, and $z$-rotation pulse time $t_z=\frac{3\pi\hbar}{h_{z,1}}(1\pm0.001)$ [with magnetic field $h_{z,1}$, Eq.~\eqref{networkhamiltonian}], see Appendix \ref{ABflux}. }
	\label{pic02} 
\end{figure}

We can use the interaction defined in Eq.~\eqref{vfdnakfvm01} to realize a controlled phase-flip  gate for two qubits. For $ \varphi = \pi\pm \alpha/2$ (for $\varphi \neq 0$ or $\pi$), we get an Ising interaction 
\begin{align}
2J \sin^2( \alpha/2)[S_1^{z} S_2^{z}\mp \left(\textstyle\frac{1}{2}S_1^{z}-\frac{1}{2}S_2^{z}\right)],
\end{align} 
where we choose our quantization $z$ axis along ${\bf u}$.
The Ising interaction enables a controlled phase-flip gate $U_{\text{CPF}}\equiv\text{diag}(1,1,1,-1)=U(\alpha,\varphi= \pi \pm \frac{\alpha}{2})$ via the evolution \cite{Loss1997}, defined for $\alpha\neq 0$, as
\begin{align}\label{eq:evolution}
U(\alpha,\varphi)=e^{-i\scriptstyle\frac{\pi}{4}} e^{i\pi S_{n(\varphi)}^z} e^{{\scriptstyle\frac{-i\pi}{J_{\Vert}(\alpha,\varphi)}}H(\alpha,\varphi)},
\end{align}  
where $n(\varphi)=1$ for $0<\varphi<\pi$ and $n(\varphi)=2$ for $\pi<\varphi<2\pi$. 
The controlled phase-flip gate is sufficient for universal quantum computing, as it is related to the CNOT gate by single-qubit rotations \cite{Loss1997} (single-qubit gates to be explained below). The advantage of using the Ising interaction instead of the isotropic interaction $\propto\mathbf{S}_1\cdot\mathbf{S}_2$ is that the latter requires two interaction gates, for example, via two square-root-of-SWAP gates \cite{Loss1997}, while the Ising interaction only requires one interaction gate. This allows faster switching times and decreases the error probability of the gate \cite{Burkard1999}. 
If we consider an aluminum superconductor with $\Delta\approx 0.1$\,meV \cite{Deacon2015} and tune the dot energy to $\epsilon\approx-10\,\mu$eV and $\Gamma \approx1\,\mu$eV, the interaction strength becomes $J\approx0.1\,\mu$eV, which corresponds to a switching time of $t=\frac{\hbar\pi}{J_{\Vert}(\alpha,\varphi)}\approx 10$\,ns for $\alpha=\pi$, $\varphi=\frac{\pi}{2}$. This time is more than 1000 times smaller than the lifetime of a trapped quasiparticle \cite{janvier2015coherent, hays2018direct, hays2021coherent}. As noted above, the interaction becomes Ising-like for $\varphi = \pi\pm \alpha/2$, which means that it is sufficient to tune either $\varphi$ or $\alpha$, but not both. 
This becomes clear when we calculate the fidelity  \cite{Huang2019}
\begin{equation}\label{eq:fidelity}
F=\left|\frac{1}{4}\textstyle\text{tr}\, U^{\dagger}_{\text{CPF}} U(\alpha,\varphi)\right|^2,
\end{equation}
where the parameters in $U(\alpha,\varphi)$ vary around their ideal values. 
In Fig.~\ref{pic02} we show the infidelity $1-F$ as a function of $\alpha$ and $\varphi$. We average over some statistical Gaussian noise of the switching time, the tunnel ratio $r=\frac{\Gamma_L}{\Gamma_R}$ (see Appendix \ref{ABflux}) and the pulse time of the single-qubit $z$ rotation. By tuning either $\alpha$ or $\varphi$, we get a fidelity $>99.99\%$.

\section{Scalable network} \label{scalablenetwork}
Based on the results from the preceding sections, we next propose a scalable network of superconducting spin qubits, see Fig.~\ref{pic01}(g). The network consists of parallel nanowires in $x$ direction and orthogonal superconducting stripes in $z$ direction.  A crossbar design \cite{Li2018} of gates confines quantum dots in the nanowires, controls the tunnel barriers between the dots and the superconductors, and allows one to address qubits individually. In summary, the total Hamiltonian of the network is
\begin{equation}\label{networkhamiltonian}
H_{\text{net}}= \sum_{ij} %
H_{ij}+ 
\sum_i (h_{x,i} S_{x,i} +h_{z,i} S_{z,i}),
\end{equation} 
where $H_{ij}(\alpha_{ij},\varphi_{ij},\Gamma_{L,ij},\Gamma_{R,ij})$ describes the interaction between dot $i$ and dot $j$ [see Eq.~\eqref{vfdnakfvm01}], and $h_{x/z,i}$ is the global/effective magnetic field of dot $i$ (including the Bohr magneton  and the corresponding $g$ factors).

Scalable single-qubit gates are implemented in the usual way as for regular spin qubits \cite{Li2018}. A small global magnetic field in $x$ direction splits the spin states. The spins can now be flipped via EDSR  \cite{Rashba2003orbital, Nowack2007,nadj2010spin,Froning2021}. An electric microwave pulse is sent to the tunnel barrier gates: due to SOI the spins experience an effective magnetic field in the $z$ direction. We assume that the global magnetic field is small enough not to affect the two-qubit gates.

For two-qubit gates there are two possibilities in our scheme, depending on the relative position of the dots. Both possibilities have in common that the superconducting spin qubits start to interact if barriers are lowered. If two neighboring quantum dots are positioned on the same nanowire [as in Fig.~\ref{pic01}(b)], they only interact through one superconductor and the interaction will be an ``ordinary twisted'' interaction. As discussed in the previous section and in Ref.~\cite{GonzalezRosado2021}, this means the interaction can be converted into an isotropic interaction by additional single-qubit rotations or by tuning the spin-orbit strength with the gates. This allows one to perform two square-root-of-SWAP gates, equivalent to a CNOT gate \cite{Loss1997,Burkard1999}. For two spins on neighboring nanowires [as in Fig.~\ref{pic01}(a)] the exchange interaction between them will be ``anomalously twisted''. As discussed in the previous section, this allows one to perform a controlled phase-flip gate, which is related to a CNOT gate via single-qubit rotations. The advantage of this interaction is that it is tunable by the superconducting phase difference and only requires one interaction gate.

The two-dimensional setup allows error correction codes such as the surface code~\cite{bravyi1998quantum}. %
Furthermore, in principle this setup allows for increased qubit connectivity: When the distance between two dots is within the superconducting coherence length, also  coupling beyond nearest neighbor is possible. Finally, undesired capacitive coupling between the quantum dots and the superconductors can be suppressed by detuning the resonant frequency of the superconductor and the qubit frequency. 

\section{Conclusion} \label{conclusion}
We have found an analytic expression of the pairwise interaction of  superconducting spin qubits in the presence of SOI. The interaction consists of Ising, Heisenberg, and DM interactions and is tunable by the superconducting phase difference, the spin-orbit parameters, and the tunnel coupling. We can realize a controlled phase-flip gate and show that a high fidelity can be reached without the requirement that both the superconducting phase difference and the SOI parameters are fine-tuned. Based on these results, we propose a scalable network of superconducting spin qubits where single-qubit rotations are performed with EDSR and two-qubit gates rely on the spin interaction we have found. 
Given the fast progress in the field  \cite{hays2021coherent,kurtossy2021andreev,matsuo2021observation, Vekris2021, Vekris2022}, we believe that our proposal is within experimental reach.

\section*{Acknowledgements}
We thank Bence Hetényi and Stefano Bosco for helpful discussions. This project has received funding from the European Union’s Horizon 2020 Research and Innovation Program under Grant Agreement No. 862046 and under Grant Agreement No. 757725 (the ERC Starting Grant). This work was supported by the Swiss National Science Foundation, NCCR QSIT, and NCCR SPIN, a National Centre of Competence (or Excellence) in Research, funded by the Swiss National Science Foundation (Grant No. 51NF40-180604).

\setcounter{section}{0}
\setcounter{equation}{0}
\appendix

\section{Details of the perturbation theory} \label{fourththeory}
In this section we show the details of the perturbation theory that lead to the spin interaction [see Eq.~\eqref{spinexchangecouplingsmv1} of the main text].
Our Hamiltonian is given by Eqs.~\eqref{eq:Hdot}, \eqref{eq:Hsc}, and \eqref{eq:H_T}.
First, we perform three subsequent basis transformations, defined by Eq.~\eqref{rotationbasis01}, which do not change the Hamiltonian of the dots or superconductors because they are assumed to be rotationally invariant.%
We also diagonalize the Hamiltonian of the superconductors via a Bogoliubov transformation ($\Tilde{c}_{j\vec{k} s}=u_k\gamma_{j\vec{k} s}+ s v_{jk}\gamma_{j\shortminus\vec{k}\shortminus s}^{\dagger}$, with $u_k=\sqrt{\frac{E_k+\xi_k}{2E_k}}$, $v_{jk}=\sqrt{\frac{E_k-\xi_k}{2E_k}}e^{-i\varphi_j}$, $E _k=\sqrt{({\xi_k})^2+\Delta^2}\,$):
\begin{align}
H_{L}=\sum_{j\vec{k}\sigma} E_k \gamma_{j\vec{k}\sigma}^{\dagger}\gamma_{j\vec{k}\sigma}.
\end{align}
Excitations from the superconducting ground state are described by Bogoliubov quasiparticles ($\gamma_{j\vec{k} s}^{\dagger}$).
The tunnel Hamiltonian becomes (with $\Tilde{d}_{1\sigma}=d_{1\sigma}$)
\begin{align}
H_T=&\sum_{\substack{jn\vec{k}s\\(j,n)\neq(L,1)}}\left[t_{jn}(u_k\gamma_{j\vec{k} s}^{\dagger} + s v_{jk}^*\gamma_{j\shortminus\vec{k}\shortminus s}) \Tilde{d}_{n s} + H.c.\right]\nonumber\\
&+ \sum_{\vec{k} s s'}\left[t_{L1}(u_k\gamma_{Lk s}^{\dagger}+s v_{Lk}^*\gamma_{L\shortminus\vec{k}\shortminus s})U_{s s'}\Tilde{d}_{1 s'} + H.c.\right].
\end{align}
We see that the total unitary basis transformation $U=U_{R1}^{\dagger}U_{R2}U_{L2}^{\dagger}U_{L1}$ now only acts on the spins of electrons that tunnel between the left superconductor and dot 1. 

We now want to calculate the spin-exchange interaction between the dots mediated by the Cooper pairs in the superconductors. For this, we evaluate the fourth-order perturbation theory contribution in the weak-coupling limit $t_{jn},\Gamma_{j}\ll \Delta,|\epsilon|$ \cite{Choi2000,probst2016signatures}, 
with $\Gamma_{j}=\pi\rho_Ft_{j1}t_{j2}$ and $\rho_F$ being the normal density of states per spin of the leads at the Fermi energy:
\begin{align}\label{eq:pertubation_theory}
H=P H_T \left(\frac{1-P}{E_0-H_0}H_T\right)^3 P.
\end{align}
Here, the unperturbed Hamiltonian is $H_0=H_L+H_D$ with the ground-state energy $E_0$. The operator $P$ projects to the spin-1/2 low-energy subspace $\{|s_1,s_2;0\rangle\}_{s_1 s_2}$,  where both dots are occupied with one electron with spins $s_1$  and $s_2$ (in the $d_{1 s}$, $\Tilde{d}_{2 s}$-basis), and both superconductors are in their ground state. In other words, we evaluate $ \langle s'_1,s'_2;0|H_{\text{eff}}|s_1,s_2;0\rangle$, by summing up 
virtual tunneling paths.
We take the perturbation theory to fourth order because it is the smallest order at which spin interaction between the two quantum dots is possible. 
The first $H_T$ will always destroy one electron on one of the dots (because in the $C_d\rightarrow\infty$ limit we do not allow double occupancy) and create one Bogoliubov quasiparticle in the superconductors. Thus the first virtual intermediate state will be $|s_1,0;\gamma_{j\vec{k} s}\rangle$ or $|0,s_2;\gamma_{j\vec{k}s}\rangle$. Next, in the limit $\Delta\gg |\epsilon|$, the electron on the other dot will tunnel to the superconductor and destroy the Bogoliubov particle again, in total creating a Cooper pair. The second virtual intermediate state is then $|0,0;0\rangle$. This process will dominate, as it costs more energy to create a second Bogoliubov quasiparticle than to remove an electron from the dot.
The third virtual intermediate state involves a Bogoliubov quasiparticle and an electron, similar to the first virtual intermediate state, and after the fourth tunneling we get back to our ground state. 

For virtual paths that involve the right superconductor only [see Fig.~\ref{pic01}(d) of the main text] one can permute the order of the first two tunneling events and of the third and fourth tunneling events, giving a total of four virtual tunneling paths. It turns out that each of them gives the same contribution to $ \langle s'_1,s'_2;0|H_{\text{eff}}|s_1,s_2;0\rangle$, namely,
\begin{align}
t_{R1}^2t_{R2}^2\sum_{\vec{k},\vec{q}}\frac{u_kv^*_{Rk}u_qv_{Rq}}{(E_k-\epsilon)2\epsilon(E_q-\epsilon)} s_1 s_1'\delta_{s_1,-s_2}\delta_{s_1',-s_2'}.
\end{align}
We evaluate
\begingroup
\allowdisplaybreaks
\begin{align}
\sum_{\vec{k},\vec{q}}&\textstyle\frac{|u_k v_{Rk} u_q v_{Rq}|}{(E_k-\epsilon)2\epsilon(E_q-\epsilon)}\\\nonumber
&={\displaystyle\frac{1}{8\epsilon}} \Bigg(\sum\limits_{\vec{k}} \textstyle \frac{1}{\sqrt{\left(\frac{\xi_k}{\Delta}\right)^2+1}\left(\sqrt{\left(\frac{\xi_k}{\Delta}\right)^2+1}-\frac{\epsilon}{\Delta}\right)\Delta }\Bigg)^2\\\nonumber
&= {\displaystyle\frac{1}{8\epsilon}}\Bigg[\int\limits_{-\Lambda}^{\Lambda} \textstyle \frac{\rho(\xi)}{ \sqrt{\left(\frac{\xi}{\Delta}\right)^2+1} \left(\sqrt{\left(\frac{\xi}{\Delta}\right)^2+1} -\frac{\epsilon}{\Delta} \right)\Delta} \text{d}\xi \Bigg] ^2\\\nonumber
&\approx \frac{\rho_F^2}{8\epsilon}\Bigg[\int\limits_{-\infty}^{\infty} \textstyle \frac{\text{d}x}{\sqrt{x^2+1}(\sqrt{x^2+1}-\frac{\epsilon}{\Delta})}\Bigg]^2\overset{\Delta\gg\epsilon}{\approx}-\displaystyle\frac{\rho_F^2\pi^2}{8|\epsilon|}.\notag
\end{align}%
\endgroup
Here, $\Lambda$ is some cut-off energy (of the order of the Fermi energy) and $\rho(\xi)$ is the density of states per spin, which we assume to be approximately constant $\rho(\xi)\approx\rho(0)=\rho_F$ in the region around the Fermi energy, which contributes most to the integral.
Further, we can replace $s_1 s_1'\delta_{s_1,-s_2}\delta_{s_1',-s_2'} \to -2\vec{S}_1\cdot\Tilde{\vec{S}}_2-\frac{1}{2}$, where the $-\frac{1}{2}$ is an irrelevant constant. The operator $\vec{S}_1$ is the spin operator of dot $n=1$ and $\Tilde{\vec{S}}_2$ is the spin operator of dot $n=2$ in the rotated basis [see Eq.~\eqref{rotationbasis01}]: $\tilde{\mathbf{S}}_2=U_{R2}U^{\dagger}_{R1}\mathbf{S}_2 U_{R1}^{\dagger}U_{R2}$.
All virtual paths corresponding to processes shown in Fig.~\ref{pic01}(d)  together give
\begin{align}
\frac{\rho_F^2\pi^2}{|\epsilon|}t_{R1}^2t_{R2}^2\vec{S}_1\cdot\Tilde{\vec{S}}_2.
\end{align}

For virtual paths involving the left superconductor only [see Fig.~\ref{pic01}(c)] we can use the result for the right lead and simply rotate the spin operator of the upper dots by  $U$, as the ``left-lead-paths'' and ``right-lead-paths'' are equivalent up to a basis rotation:
\begin{align}
\frac{\rho_F^2\pi^2}{|\epsilon|}t_{L1}^2t_{L2}^2(U^{\dagger} \vec{S}_1 U)\cdot\Tilde{\vec{S}}_2.
\end{align}

For virtual paths where a Cooper pair transfers from the left to the right superconductor [see Fig.~\ref{pic01}(e)],  there exist four permutations of tunneling paths. Each of them contributes with
\begin{align}\label{eq:left-to-right-contr}
-\frac{\rho_F^2\pi^2}{8|\epsilon|}t_{L1}t_{L2}t_{R1}t_{R2}e^{-i\varphi}(-1)s_1 s_2'\delta_{s_1-s_2}U^*_{\shortminus s_2' s_1'}.
\end{align}
Using that 
\begin{align}
U&=\textstyle\cos(\frac{\alpha}{2})+2 i\vec{u}\cdot\vec{S}_1\sin(\frac{\alpha}{2})\\\nonumber
&=\begin{pmatrix}\cos{(\frac{\alpha}{2})}+iu_z\sin{(\frac{\alpha}{2})}&(u_y+iu_x)\sin{(\frac{\alpha}{2})}\\(-u_y+iu_x)\sin{(\frac{\alpha}{2})}&\cos{(\frac{\alpha}{2})}-iu_z\sin{(\frac{\alpha}{2})}\end{pmatrix},
\end{align}
we simplify 
\begin{widetext}
\begin{align}
s_1 s_2'\delta_{s_2-s_1}U^*_{\shortminus s_2' s_1'}  \to \,\,
&\textstyle 2\cos(\frac{\alpha}{2})(\vec{S}_1\cdot\Tilde{\vec{S}}_2-\frac{1}{4})
+u_z\sin(\frac{\alpha}{2})(2S_{1y}\Tilde{S}_{2x}-2S_{1x}\Tilde{S}_{2y}+iS_{1z}-i\Tilde{S}_{2z}) \\\nonumber
&+u_y\sin\textstyle(\frac{\alpha}{2})(2S_{1x}\Tilde{S}_{2z}-2S_{1z}\Tilde{S}_{2x}+iS_{1y}-i\Tilde{S}_{2y}) 
+u_x\sin(\frac{\alpha}{2})(2S_{1z}\Tilde{S}_{2y}-2S_{1y}\Tilde{S}_{2z}+iS_{1x}-i\Tilde{S}_{2x}) .
\end{align}
When we reverse the tunnel sequence such that Cooper pairs travel from the right to the left superconductor, this adds the Hermitian conjugate to the expression above [Eq.~\eqref{eq:left-to-right-contr}]. Multiplying with the permutation factor 4 we get
\begin{align}
\frac{2\rho_F^2\pi^2}{|\epsilon|} t_{L1}t_{L2}t_{R1}t_{R2}\Bigg\{\!\cos{\varphi}
\Big[\!\cos\textstyle(\frac{\alpha}{2})&\vec{S}_1\!\cdot\Tilde{\vec{S}}_2\!-\!\sin(\textstyle\frac{\alpha}{2})\vec{u}\cdot\!(\vec{S}_1\!\times\!\Tilde{\vec{S}}_2)\Big]
+\frac{1}{2}\sin(\varphi)\sin(\textstyle\frac{\alpha}{2})\vec{u}\cdot\!(\vec{S}_1\!-\!\Tilde{\vec{S}}_2)\Bigg\}+g(\varphi),\\\nonumber
&g(\varphi)=-\frac{\rho_F^2\pi^2}{2|\epsilon|}t_{L1}t_{L2}t_{R1}t_{R2}\cos{\varphi}\cos{\left(\frac{\alpha}{2}\right)}.
\end{align} 

Adding all contributions together, we calculate the effective Hamiltonian in the spin-1/2 subspace to be 
\begin{align}
H=&
\frac{\rho_F^2\pi^2}{|\epsilon|}\Bigg(t_{R1}^2t_{R2}^2\vec{S}_1\cdot\Tilde{\vec{S}}_2 +t_{L1}^2t_{L2}^2 (U^{\dagger}\vec{S}_1 U) \cdot\Tilde{\vec{S}}_2\\\nonumber
&+ 2t_{L1}t_{L2}t_{R1}t_{R2}\Big\{\!\cos{\varphi}
\Big[\!\cos(\textstyle\frac{\alpha}{2})\vec{S}_1\!\cdot\Tilde{\vec{S}}_2\!-\!\sin(\textstyle\frac{\alpha}{2})\vec{u}\cdot\!(\vec{S}_1\!\times\!\Tilde{\vec{S}}_2)\Big] 
+\frac{1}{2}\sin(\varphi)\sin(\textstyle\frac{\alpha}{2})\vec{u}\cdot\!(\vec{S}_1\!-\!\Tilde{\vec{S}}_2)\Big\}\Bigg)+g(\varphi).
\end{align}
\end{widetext}

This equation is Eq.~\eqref{spinexchangecouplingsmv1} of the main text for $t_{L1}t_{L2}=t_{R1}t_{R2}$.

We note that the spin Hamiltonian in Eq.~\eqref{vfdnakfvm01} can be viewed as a form of Ruderman-Kittel-Kasuya-Yosida (RKKY) interaction  \cite{ruderman1954indirect,kasuya1956theory,yosida1957magnetic} between the dot spins which is mediated by Cooper pairs that virtually split and recombine or vice versa.

\section{Ahoronov-Bohm flux} \label{ABflux}
In this section we present a more general result of the effective Hamiltonian found by using the perturbation  theory {\it without} the restriction $\Delta\gg|\epsilon|$ and accounting for an Aharonov-Bohm flux $f$ inclosed by the two quantum dots and two superconductors. We note that in the limit $\Delta\gg|\epsilon|$, those terms that depend on the Aharonov-Bohm flux are higher-order contributions \cite{Choi2000} and correspond to virtual tunneling paths as depicted in Fig.~\ref{pic01}(f) of the main text.

We adopt a more general description of the tunnel Hamiltonian, which is $\vec{k}$ dependent and includes a Peierls phase in the presence of an external magnetic vector potential $\vec{A}$. For this we replace $t_{jn}$ by $t_{jn}\exp\left(\shortminus i\vec{k}\cdot\vec{r}_j-{\textstyle \frac{i\pi}{\phi_0}}\int\limits_{\vec{r}_n}^{\vec{r}_j}\!\text{d}\vec{l}\cdot\!\vec{A}\right)$, where $\vec{r}_j$ is the position on the superconductor that couples to both quantum dots, each located at position $\vec{r}_n$. In addition, $\phi_0=\frac{hc}{2e}$ is the superconducting flux quantum. (We assume the associated Zeeman splitting on the dot to be small and neglect it.) From the symmetry arguments presented in the main part, it still follows that the total Hamiltonian has a similar form as Eq.~\eqref{spinexchangecouplingsmv1} in the main text. The generalized result then becomes ({\it without} the restriction $\Delta\gg|\epsilon|$):
\begin{widetext}
\begin{align} \label{fvnkdv}
H&=\textstyle\Gamma_{R}^2(\frac{C}{|\epsilon|}+\frac{C_0+C_1}{\Delta})\mathbf{S}_1\cdot\Tilde{\mathbf{S}}_2+\Gamma_{L}^2(\frac{C}{|\epsilon|}+\frac{C_0+C_1}{\Delta})(U^{\dagger}\mathbf{S}_1 U)\cdot\Tilde{\mathbf{S}}_2\\
&\textstyle+ 2\Gamma_L\Gamma_R\left[\frac{C_1}{\Delta}\cos (f)+(\frac{C}{|\epsilon|}+\frac{C_0}{\Delta})\cos(\varphi)\right]
\left[\cos(\frac{\alpha}{2})\mathbf{S}_1\cdot\Tilde{\mathbf{S}}_2-\sin(\frac{\alpha}{2})\mathbf{u}\cdot(\mathbf{S}_1\times\Tilde{\mathbf{S}}_2)\right]\notag\\
&\textstyle+\Gamma_L\Gamma_R\sin\left(\frac{\alpha}{2}\right)\left[\frac{C_1}{\Delta}\sin(f)\mathbf{u}\cdot(\mathbf{S}_1+\Tilde{\mathbf{S}}_2)+(\frac{C}{|\epsilon|}+\frac{C_0}{\Delta})\sin(\varphi)\mathbf{u}\cdot(\mathbf{S}_1-\Tilde{\mathbf{S}}_2)
\right]+g(\varphi,f).  \notag 
\end{align}
In order to reveal the structure of the spin interaction, we rewrite Eq.~\eqref{fvnkdv} in a  form similar to Eq.  \eqref{vfdnakfvm01} in the main text: 
\begin{equation} \label{vysfdnakfvm01}
H=h(S^{\Vert}_{1}-\tilde{S}^{\Vert}_{2})+h_z(S^{\Vert}_{1}+\tilde{S}^{\Vert}_{2})+J_{\Vert} S^{\Vert}_{1}\tilde{S}^{\Vert}_{2}+ J_{\perp} \mathbf{S}^{\perp}_{1}\cdot \tilde{\mathbf{S}}^{\perp}_{2}+J_{\text{DM}}(\mathbf{S}_1\times\tilde{\mathbf{S}}_2)^{\Vert}+g(\varphi,f).
\end{equation}
Here,  the effective magnetic fields, spin-exchange constant, and a spin-irrelevant constant $g(\varphi,f)$, respectively, are given by
\begin{subequations}
\begin{align}
&h=\textstyle\Gamma_L\Gamma_R\left(\frac{C}{|\epsilon|}+\frac{C_0}{\Delta}\right)\sin\left(\frac{\alpha}{2}\right)\sin(\varphi), \\
&h_z=\textstyle\Gamma_L\Gamma_R\frac{C_1}{\Delta}\sin\left(\frac{\alpha}{2}\right)\sin(f), \\
&J_{\Vert}= \textstyle(\Gamma_{R}^2+\Gamma_{L}^2)\left(\frac{C}{|\epsilon|}+\frac{C_0+C_1}{\Delta}\right)+2\Gamma_L\Gamma_R\left[\frac{C_1}{\Delta}\cos (f)+\left(\frac{C}{|\epsilon|}+\frac{C_0}{\Delta}\right)\cos(\varphi)\right]
\cos(\frac{\alpha}{2}),
\\ &J_{\perp}=\textstyle\left[\Gamma_{R}^2+\Gamma_{L}^2\cos\left(\alpha\right)\right]\left(\frac{C}{|\epsilon|}+\frac{C_0+C_1}{\Delta}\right)+2\Gamma_L\Gamma_R\left[\frac{C_1}{\Delta}\cos (f)+\left(\frac{C}{|\epsilon|}+\frac{C_0}{\Delta}\right)\cos(\varphi)\right]
\cos(\frac{\alpha}{2}),\\\ 
&J_{\text{DM}}=-\textstyle\Gamma_{L}^2\left(\frac{C}{|\epsilon|}+\frac{C_0+C_1}{\Delta}\right)\sin(\alpha)-2\Gamma_L\Gamma_R\left[\frac{C_1}{\Delta}\cos (f)+\left(\frac{C}{|\epsilon|}+\frac{C_0}{\Delta}\right)\cos(\varphi)\right]
\sin(\frac{\alpha}{2}),\\
&g(\varphi,f)=\textstyle\frac{1}{2}\Gamma_L\Gamma_R\left[\frac{C_1}{\Delta}\cos(f)\cos(\frac{\alpha}{2})-(\frac{C_0}{\Delta}+\frac{C}{|\epsilon|})\cos(\varphi)\cos(\frac{\alpha}{2})\right]+(t_{L1}^2t_{R1}^2+t_{L2}^2t_{R2}^2)\frac{\rho_F^2\pi^2C_0}{2\Delta}\cos(\varphi).
\end{align}
\end{subequations}
\end{widetext}
We have parameterized the tunneling coupling by $\Gamma_j=\rho_F\pi t_{j1}t_{j2}$. The dimensionless parameters $C$, $C_0$, and $C_1$ are given by 
\begin{align}
C&=\left[\int\limits_{-\infty}\limits^{\infty}\frac{\text{d}x}{\pi h_0(x)g_0(x)}\right]^2\\
C_0&=\int\limits_{-\infty}\limits^{\infty}\int\limits_{-\infty}\limits^{\infty}\frac{\text{d}x\,\text{d}y}{\pi^2 h_0(x)h_0(y)g_0(x)g_0(y)[h_0(x)+h_0(y)]}\\
C_1&=\int\limits_{-\infty}\limits^{\infty}\int\limits_{-\infty}\limits^{\infty}\frac{\text{d}x\,\text{d}y\,[h_0(x)-\frac{|\epsilon|}{\Delta}]}{\pi^2[h_0(x)+h_0(y)][g_0(x)]^2g_0(y)},
\end{align}
where $h_0(x)=\sqrt{x^2+1}$, $g_0(x)=\sqrt{x^2+1}+\frac{|\epsilon|}{\Delta}$ \cite{Choi2000}.
The phase $\varphi$ and Aharonov-Bohm flux $f$, respectively, are given by 
\begin{align}
    \varphi=\varphi_{L}-\varphi_R-\frac{\pi}{\phi_{0}} \int_{\mathbf{r}_{R}}^{\mathbf{r}_{L}}(d \vec{\ell}_{1} +d \vec{\ell}_{2})\cdot \mathbf{A},
\end{align}
\begin{align}
    f=\frac{\pi}{\phi_{0}} \int_{\mathbf{r}_{R}}^{\mathbf{r}_{L}}(d \vec{\ell}_{1} - d \vec{\ell}_{2})\cdot \mathbf{A}.
\end{align}
Here $\vec{\ell}_{n}$ corresponds to the path $\vec{r}_{R}\rightarrow\vec{r}_{n}\rightarrow\vec{r}_{L}$ 
and $f$ is the dimensionless Aharonov-Bohm flux $f$ running through the area between the dots and the superconductors \cite{Choi2000}.
In addition to the effective staggered magnetic fields [first term in Eq.~\eqref{vysfdnakfvm01}] there exists a symmetry-allowed Zeeman term [second term in Eq.~\eqref{vysfdnakfvm01}]. Generally, the external flux $f$ offers one more experimentally tunable parameter, which increases the tunability of our scalable architecture. Again, if $\Gamma_L= \Gamma_R$, we can achieve the purely Ising coupling with $J_\perp=J_{DM}=0$: 
\begin{align}\label{is}
&\left(\frac{C}{|\epsilon|}+\frac{C_0+C_1}{\Delta}\right)\cos\left(\frac{\alpha}{2}\right)=\\\nonumber 
&-\left[\frac{C_1}{\Delta}\cos (f)+\left(\frac{C}{|\epsilon|}+\frac{C_0}{\Delta}\right)\cos(\varphi)\right].
\end{align}
In the absence of flux, Eq. (\ref{is}) reproduces the criterion derived in the main part. The flux $f$ gives us one more parameter to be used to achieve the optimal regime with high fidelities (see Fig.~\ref{pic02}). We also note that, if $\Gamma_L\neq \Gamma_R$ it is not possible to achieve the pure Ising regime.

\section{Josephson supercurrent} \label{Josephonsupercurrent}
The detection of spin interactions can be also realized by observing the Josephson supercurrent $\hat I$, being defined by the derivative of $H$ with respect to $\varphi$, i.e., $\hat I=\partial H/\partial \varphi$. Using Eq.~\eqref{vysfdnakfvm01}, we obtain for the  supercurrent  
\begin{align} \label{dfvmkfmvk}
    \hat{I}={\hat I}_s\sin\varphi+I_0\sin\left(\frac{\alpha}{2}\right)(S^{\Vert}_{1}-\tilde{S}^{\Vert}_{2})\cos\varphi,
\end{align}
with 
\begin{align} \label{fdjvo2}
\hat I_s=-2I_0
\Big[
&\cos\left(\frac{\alpha}{2}\right)\mathbf{S}^{}_{1}\cdot \tilde{\mathbf{S}}^{}_{2}-\sin\left(\frac{\alpha}{2}\right)(\mathbf{S}_1\times\tilde{\mathbf{S}}_2)^{\Vert}\\\nonumber
&-\frac{1}{4}\cos\left(\frac{\alpha}{2}\right)\Big]-I_1.
\end{align}
Note that the Josephson current $\hat I$ is still an operator in spin space, indicated by the hat.
This current involves  only ``two-lead'' paths, as manifested by the current amplitudes that contain simultaneously  tunneling amplitudes from right and left superconducting leads:
\begin{align}
    I_{0}=\Gamma_L\Gamma_R\left(\frac{C}{|\epsilon|}+\frac{C_0}{\Delta}\right),
\end{align}
\begin{align}
    I_{1}=(t_{L1}^2t_{R1}^2+t_{L2}^2t_{R2}^2)\frac{\rho_F^2\pi^2C_0}{2\Delta}.
\end{align}
In the presence of SOI ($\alpha\neq0$), the supercurrent   contains an anomalous term proportional to $\cos \varphi$. This means that the supercurrent $\hat I$ is finite even if  $\varphi=0$. This phase shift in the supercurrent can be exploited to detect the presence of the SOI, depending on the orientations of the dot spins relative to each other. In turn, this spin dependence of the phase shift can also be used as a readout of the spin states of the dots: if one spin is aligned  along its quantization axis, while the other one is antialigned along its respective quantization axis, the amplitude of the Josephson current at $\varphi=0$ will be maximal for finite SOI. In the opposite extreme case, when the spins are both aligned along their respective quantization axes, the current is minimal  at $\varphi=0$, i.e., it vanishes.
 
For the special case considered in the main text, $\Gamma_L=\Gamma_R=\Gamma$ and $\vert \epsilon\vert \ll\Delta$, we have $C_0\simeq 0$ and $C\simeq 1$ (and $C/\vert \epsilon\vert \gg C_0/\Delta)$. Then Eqs.~\eqref{dfvmkfmvk} and \eqref{fdjvo2} reduce to Eqs.~\eqref{1230987123987} and \eqref{454343450908}, respectively, of the main text.

\twocolumngrid

\end{document}